\documentclass[aps,prd,amssymb,nofootinbib,twocolumn,epsf,floatfix,showpacs]
{revtex4}
\usepackage[usenames]{color} 
\usepackage{ulem} 
\usepackage{graphicx}
\usepackage{amssymb}
\usepackage{amsmath}
\usepackage{epstopdf}

\begin{document}

%%%%%%%%%%%%%%%%%%%%%%%%%%%%%%%%%%%%%%%%%%%%%%%%%%%%%%%%%%%%%%%%%%%%%%%%%%%%%%%

\title{Model Waveform Accuracy Requirements for the Allen $\chi^2$ 
Discriminator}

\author{Lee Lindblom${}^{1}$ and Curt Cutler${}^{\,2,3}$} 

\affiliation{${}^1$Center for Astrophysics and Space Sciences, University 
of California at San Diego, 9500 Gilman Drive, La Jolla, CA 92093, USA}

\affiliation{${}^2$ Theoretical Astrophysics 350-17, California Institute of
Technology, Pasadena, California 91125, USA}

\affiliation{${}^3$Jet Propulsion Laboratory, M/S 169-327, 4800 Oak Grove Drive, Pasadena, CA 91109, USA }

\date{\today}
 
\begin{abstract} 
This paper derives accuracy standards for model gravitational
waveforms required to ensure proper use of the Allen $\chi^2$
discriminator in gravitational wave (GW) data analysis.  These
standards are different from previously established requirements for
detection and waveform parameter measurement based on signal-to-noise
optimization.  We present convenient formulae for evaluating and
interpreting the contribution of model errors to measured values of
this $\chi^2$ statistic.  The new accuracy standards derived here are
needed to ensure the reliability of measured values of the Allen
$\chi^2$ statistic, both in their traditional role as vetoes and in
their current role as elements in evaluating the significance of
candidate detections.
\end{abstract}
 
 
\maketitle

%%%%%%%%%%%%%%%%%%%%%%%%%%%%%%%%%%%%%%%%%%%%%%%%%%%%%%%%%%%%%%%%%%%%%%%%%%%%%%%
\section{Introduction}
\label{s:Introduction}
%%%%%%%%%%%%%%%%%%%%%%%%%%%%%%%%%%%%%%%%%%%%%%%%%%%%%%%%%%%%%%%%%%%%%%%%%%%%%%%

For most potential astrophysical sources of gravitational waves (GWs),
including the orbital inspiral and merger of two black holes, the
exact solutions of Einstein's equations that describe them are not
known.  Therefore matched-filter searches for the GWs emitted by these
systems must rely on approximate model waveforms.  Standard
approximation methods include: the post-Newtonian approximation, the
effective one-body approximation, the large-mass-ratio approximation,
and numerical relativity.  What are the requirements on the accuracy
of these approximate gravitational waveforms set by the practical
needs of GW data analysts?  In previous work, one of us (LL) examined
in detail how waveform inaccuracy impacts signal-to-noise ratios
(SNRs)~\cite{Lindblom2008}, and derived sufficient conditions on
waveform accuracy to ensure that detection rates and waveform
parameter measurements are not significantly affected by waveform
errors.  In addition one of us (CC) developed formulae relating the
systematic errors in the inferred physical parameters of a binary
inspiral waveform (e.g., the masses of the two bodies) to the model
errors in the waveform~\cite{Cutler_Vallisneri_2007}.  For both
detection and parameter-estimation purposes, a reasonable goal for
theoretical waveform modelers is to insure that errors (e.g., false
dismissals or parameter-estimation errors) due to the intrinsic
detector noise dominate over errors due to inaccurate waveform models.

In this paper we consider the requirements on waveform accuracy needed
for the use of the Allen $\chi^2$ discriminator in GW data
analysis.  This $\chi^2$ discriminator was introduced
by Bruce Allen~\cite{Allen2005} to provide a veto against instrumental
glitches in GW detectors that, because of their large amplitude, give
a high matched-filter SNR value, but which do not actually resemble
the waveforms used as search templates.  This
$\chi^2$ discriminator measures how well the frequency domain
structure of a putative GW signal agrees with the frequency-domain
structure of the model waveform used to detect it.  In current LIGO
data analysis this Allen $\chi^2$ discriminator is {\it not}\,
used by itself as a veto on candidate GW signals, i.e., there is no
threshold value of this $\chi^2$ such that candidates with
higher values are simply discarded.  Instead, the Allen $\chi^2$
statistic is now used along with the standard matched-filter SNR to
produce a {\it re-weighted}\, SNR that is used to assess the
statistical significance of candidate
detections~\cite{FirstDiscovery}.  This re-weighted SNR is more
effective for estimating this significance in the presence of
realistic non-Gaussian noise in the data than the standard
SNR~\cite{Allen2005, BabakEtAl2013}.

The accuracy standards derived here are quite general, requiring only
that errors in the model waveform have no more effect on the Allen
$\chi^2$ discriminator than statistical noise in the detector.  These
standards are important to insure that the values of this $\chi^2$
statistic currently used in the assessment of the significance of
candidate signals are reliable.  And we feel that these waveform
accuracy requirements will continue to be relevant for future uses of
the Allen $\chi^2$ discriminator in GW searches for the following
reasons.  i) In the future other uses of this $\chi^2$ statistic in GW
data analysis may be developed in which the influence of model errors
could be even more important.  ii) We have derived a number of simple,
convenient formulae describing how model waveform error affects the
Allen $\chi^2$ statistic, and these could be useful in developing
future data analysis applications.  iii) It has often been suggested
that (some version of) the Allen $\chi^2$ statistic could be used for
model verification (e.g., is general relativity the correct theory of
gravitation, and is the observed inspiral waveform actually produced
by black holes as opposed to some more exotic type of compact object
like a boson star)~\cite{LIGO_GR_TESTS_2016}.  iv) The accuracy
standards associated with the Allen $\chi^2$ statistic should be
useful to waveform modelers right now by providing a new, simple
figure of merit for assessing model waveform accuracy.

The remainder of this paper is organized as follows.  In
Sec.~\ref{s:Review} we briefly review relevant basic background
material on GW data analysis.  In Sec.~\ref{s:WaveformAccuracy} we
derive expressions describing how model gravitational waveform errors
affect measured values of the Allen $\chi^2$ statistic, and then
derive our Allen $\chi^2$ discriminator-based requirement on waveform
accuracy.

%%%%%%%%%%%%%%%%%%%%%%%%%%%%%%%%%%%%%%%%%%%%%%%%%%%%%%%%%%%%%%%%%%%%%%%%%%%%%%%
\section{GW Data Analysis: Background}
\label{s:Review}
%%%%%%%%%%%%%%%%%%%%%%%%%%%%%%%%%%%%%%%%%%%%%%%%%%%%%%%%%%%%%%%%%%%%%%%%%%%%%%%
This section contains short summaries of some relevant background
material on GW data analysis: matched-filter methods for GW searches,
the Allen $\chi^2$ discriminator, and previous work on how
inaccuracies in model gravitational waveforms impact GW data analysis.
A more comprehensive discussion of gravitational wave data analysis
can be found in many references, including for example Creighton and
Anderson~\cite{Creighton2011} and references cited therein.  An
up-to-date summary of how data are being analyzed in the first
advanced LIGO observing runs is given in Abbott et
al.~\cite{FirstResults}.

\subsection{Matched-filter searches}
Let $h_e(t,{\mathbf \lambda_e})$ denote the exact gravitational
waveform from a particular astrophysical source with physical
parameters ${\mathbf \lambda_e}$.  It is most convenient to describe
the matched-filter approach to GW data analysis in terms of the
Fourier transforms of the waveforms.  Let $h_e(f,\mathbf \lambda_e)$
denote the Fourier transform of the exact waveform:
\begin{eqnarray}
h_e(f,\mathbf \lambda_e)=\int^\infty_{-\infty} h_e(t,\mathbf
\lambda_e) e^{-2\pi i f t} dt.
\label{e:FourierTransformDef}
\end{eqnarray}
Signals are detected in the noisy output data stream from a GW
detector by searching for model waveforms $h_m(f,{\mathbf \lambda_m})$
that provide a sufficiently good match to the exact waveform of the
signal embedded in that data.  This matching is done by
projecting the Fourier transforms of model waveforms onto the GW
signal using the noise weighted (complex) inner product $\langle
h_e|h_m\rangle$, defined by
\begin{eqnarray}
\langle h_e|h_m\rangle=4\int_0^\infty \frac{h_e(f)h^*_m(f)}
{S_n(f)}df,
\label{e:InnerProductDef}
\end{eqnarray}
where $S_n(f)$ is the one-sided power spectral density of the detector
strain noise.

Matched-filter searches for GW signals begin by looking
for a model waveform $h_m({\mathbf \lambda_m})$ that agrees with the
signal $h_e(\mathbf \lambda_e)$ to some level of accuracy.  One
measure of this agreement is the signal-to-noise ratio (SNR),
$\rho_m({\mathbf \lambda_m})$, defined by
\begin{eqnarray}
\rho_m^{\,2}({\lambda_m})=\frac{\Bigl|\langle h_e| h_m({\mathbf
    \lambda_m})\rangle\Bigr|^2} {\langle h_m({\mathbf
    \lambda_m})|h_m({\mathbf \lambda_m})\rangle}.
\label{e:SignalToNoiseDef}
\end{eqnarray}
The quantity $\rho_m$ measures the projection of the signal $h_e$ onto
the model waveform $h_m$, using the noise-weighted inner product given
in Eq.~(\ref{e:InnerProductDef}).  Thus $\rho_m$ measures the
component of $h_e$ described by the model waveform $h_m$ in units of
the noise level of the detector.  The best fit waveform model for a
particular signal $h_e$ is obtained by adjusting the model parameters
${\mathbf \lambda_m}$ to maximize $\rho_m$.  In LIGO GW searches using
matched-filter methods, candidate signals are required to meet some
minimal threshold for $\rho_m$ (in each of at least two detectors).
This minimal detection threshold has been set at $\rho_m\gtrsim 5.5$,
for example, in recent initial LIGO searches for compact binary
signals~\cite{LIGOS6Binary} as well as the current advanced LIGO GW
searches using these methods, i.e., the PyCBC
analysis~\cite{FirstDiscovery,FirstResults}.

The parameters $\mathbf \lambda_m$ include some that represent the
intrinsic physical characteristics of the gravitational wave source (e.g., the
masses and spins of the black holes in a compact binary system), plus
extrinsic parameters, such as the relative orientations of the source and
the detector.  The model waveform $h_m$ can also be multiplied by an
arbitrary complex scale factor without changing the measured SNR
defined in Eq.~(\ref{e:SignalToNoiseDef}).  This complex scale can be
written as a real amplitude $A_0$ and phase $\phi_0$: $A_0e^{i\phi_0}$.
We are free to choose these scale parameters in any way we wish.  Here it
is convenient to fix the amplitude $A_0$ so that the model waveform
has the same overall scale as the observed signal $h_e$ by requiring
\begin{eqnarray}
\rho_m^{\,2}=\langle h_m|h_m \rangle. 
\label{e:WaveformNormalization}
\end{eqnarray}
Similarly, it is convenient to fix the phase parameter $\phi_0$ by
requiring that it match the complex phase of the observed signal by
requiring
\begin{eqnarray}
\langle h_e | h_m\rangle = \langle h_m | h_e\rangle.
\label{e:WaveformPhaseMatching}
\end{eqnarray}
We will assume in the analysis that follows that these model waveform
scale parameters have been chosen in this way according to
Eqs.~(\ref{e:WaveformNormalization}) and (\ref{e:WaveformPhaseMatching}).

\subsection{The Allen $\chi^2$ discriminator}

Allen~\cite{Allen2005} was the first to propose using the $\chi^2$
discriminator in GW data analysis.  Allen's $\chi^2$ statistic
measures how well the frequency dependence of a detected signal agrees
with that of the model waveform used to detect it.  Once a candidate
signal is identified whose measured SNR $\rho_m$ exceeds some minimal
detection threshold, the optimal model waveform $h_m$, normalized
using Eqs.~(\ref{e:WaveformNormalization}) and
(\ref{e:WaveformPhaseMatching}), is written as a sum of $p$ mutually
orthogonal components, $h_m=\sum_{k=1}^p h^k_m$.  Each component
waveform has support only in the frequency range, $f_{i-1}\leq f \leq
f_i$, chosen so that $\langle h_m^k| h_m^k\rangle = \langle
h_m|h_m\rangle/p$.  The (re-normalized) root-mean-square deviation,
$\chi^2_r$, of these component signal-to-noise quantities from their
expected values is given by
\begin{eqnarray}
\chi^2_r = \frac{p}{2p-2} \frac{1}{\rho_m^{\,2}}
\sum_{k=1}^p \left|
\langle h_e|h^k_m\rangle
- \frac{\langle h_e|h_m\rangle}{p}\right|^2.
\label{e:ChiSquareDef}
\end{eqnarray}
The expectation value and standard variation of the quantity $\chi^2_r$
(assuming stationary Gaussian detector noise) are given by the
standard expressions for a system having $2p-2$ degrees of freedom
(cf. Allen~\cite{Allen2005}),
\begin{eqnarray}
\langle \chi^2_r \rangle =1 \pm \frac{1}{\sqrt{p-1}}.
\label{e:ChiSquareRange}
\end{eqnarray}
The expressions given here are written for an arbitrary number of
frequency bins $p$.  The choice $p=16$ was typical in initial LIGO
searches (e.g., see Ref.~\cite{Canton2014}), while choosing p in a way
that depends on the properties of the waveform model, like
$p=[0.4(f_\mathrm{peak}/\mathrm{Hz})^{2.3}]$, is also being used in
advanced LIGO searches~\cite{FirstResults}.

Allen's original idea was to use the $\chi^2$ discriminator to veto
candidate signals having $\chi^2_r > \chi^2_\mathrm{th}$, for some
appropriately chosen threshold $\chi^2_\mathrm{th}$.  It was used
effectively in this way, for example, to reject large non-Gaussian
noise glitches in the analysis of the initial LIGO S5
data~\cite{BabakEtAl2013}.  The Allen $\chi^2$ discriminator
continues to play a role in GW data analysis, but its use now is less
direct.  Candidate signals having sufficiently large SNR $\rho_m$ must
now satisfy several criteria before they are considered true
gravitational wave events.  One of these criteria is a significance
test that estimates the probability the optimal model waveform $h_m$
also matches detector noise alone.  The significance of a candidate
event is determined by comparing its measured re-weighted SNR $\hat
\rho_m$ (defined below) to those obtained from a very large number of
detector noise samples.  (For the purpose of this test, the detector
noise is simulated using time shifted data from the detector.)  This
re-weighted SNR $\hat \rho_m$ reduces the standard SNR $\rho_m$ for
events having larger than expected values of $\chi^2_r$:
\begin{equation}
\hat \rho_m =\left\{\begin{array}{lr}
\rho_m  &  \mathrm{if}\,\, \chi^2_r \le 1, \\
\rho_m\bigg/\Bigl[\frac{1}{2}\Bigl(1+\bigl(\chi^2_r\bigr)^3\Bigr)
\Bigr]^{1/6}\qquad
&\mathrm{if}\,\,\chi_r^2 > 1.  
\end{array}\right. 
\end{equation}
It therefore serves as a filter that can effectively remove large
non-Gaussian noise glitches by substantially reducing their effective
SNR, but it does this in a softer way than using $\chi^2_r$ as a
strict veto.

%%%%%%%%%%%%%%%%%%%%%%%%%%%%%%%%%%%%%%%%%%%%%%%%%%%%%%%%%%%%%%%%%%%%%%%%%%%%%%%
\subsection{Waveform Accuracy and False Dismissal Rates}
\label{s:AccuracyRequirementsChiSquareTest}
%%%%%%%%%%%%%%%%%%%%%%%%%%%%%%%%%%%%%%%%%%%%%%%%%%%%%%%%%%%%%%%%%%%%%%%%%%%%%%%

In this section we review the impact of model waveform errors on false
dismissal rates.  The best-fit model waveform, $h_m$, will differ from
the exact, $h_e$, by an amount $\delta h=h_m-h_e$ that represents an
error in the model waveform.  These errors may arise either from
errors in the model waveform parameters $\lambda_m$, or from intrinsic
errors in the model waveform itself (e.g., errors from the numerical
relativity code used to produce it).  The largest SNR that could be
achieved in the absence of any model waveform error ($\delta h=0$) is
the optimal SNR $\rho_o=\langle h_e|h_e\rangle^{1/2}$.  Gravitational
wave searches using matched filter methods will miss some fraction of
the real signals unless the measured SNR $\rho_m$ is close to the
optimal $\rho_o$.  It is straightforward to determine how $\rho_m$
depends on the waveform error $\delta h$:
\begin{eqnarray}
\rho_m^{\,2}=\rho_o^{\,2}
\left[1-\frac{\langle\delta h|\delta h\rangle}{\langle h_m|h_m\rangle}
+\mathcal{O}\left(\delta h^4\right)\right],
\label{e:OptimalSNRRelation}
\end{eqnarray}
where we have assumed the waveform error $\delta h$ is small in the
sense that $|\delta h|\ll |h_m|$.  This expression uses the
fact that 
\begin{eqnarray}
0=\langle \delta h| h_m\rangle,
\end{eqnarray}
which follows as a consequence of the model-waveform scale-factor
normalization conditions in Eqs.~(\ref{e:WaveformNormalization}) and
(\ref{e:WaveformPhaseMatching}).  If follows from
Eq.~(\ref{e:OptimalSNRRelation}) that model waveform errors must be
limited by
\begin{eqnarray}
\frac{\langle\delta h | \delta h\rangle} 
{\langle h_m | h_m\rangle}
< 2\epsilon_{\mathrm{max}},
\label{e:DetectionAccuracy}
\end{eqnarray}
for some $\epsilon_\mathrm{max}$ to ensure that the measured SNR
$\rho_m$ does not differ significantly from the optimal $\rho_o$.
This result was derived in Ref.~\cite{Lindblom2008}.  We note that
while the complex inner product used in this paper is different from
the real inner product Eq.~(3) of Ref.~\cite{Lindblom2008}, the
criterion above is actually the same, since both the numerator and
denominator in Eq.~(\ref{e:DetectionAccuracy}) are real.

Previous studies~\cite{Lindblom2008} have shown that the parameter
$\epsilon_{\mathrm{max}}$ determines the fraction of real signals that
would be missed in GW searches.  The exact requirement on the value of
the parameter $\epsilon_\mathrm{max}$ that appears in
Eq.~(\ref{e:DetectionAccuracy}) is determined by the false dismissal
rate that will be tolerated in a particular search, and the details of
the data analysis procedure being used.  If we assume the
model-waveform parameters $\lambda_m$ have been adjusted to give the
optimal fit to the observed signal, then the errors in the model
waveform, $\delta h$, must be limited using
Eq.~(\ref{e:DetectionAccuracy}) with $\epsilon_\mathrm{max}=0.035$ to
ensure that no more than about 10\% of real signals are
missed~\cite{Lindblom2008}.

In actual matched-filter searches for GWs from compact binary systems,
the model-waveform parameters $\lambda_m$ are usually limited to a
discrete grid of points.  It is the combination of grid-spacing errors
and intrinsic model waveform errors that determine the false dismissal
rate.  The intrinsic waveform accuracy requirement for searches that
use discrete grids and a $10\%$ false dismissal probability must
therefore be even more stringent than $\epsilon_\mathrm{max}=0.035$.
For typical LIGO template bank searches where the maximum missmatch
between waveforms in the template bank is $\epsilon_{MM}=0.03$, the
appropriate value for $\epsilon_\mathrm{max}$ is
$\epsilon_\mathrm{max}=0.005$; see Ref.~\cite{Lindblom2008} for more
details. 

%%%%%%%%%%%%%%%%%%%%%%%%%%%%%%%%%%%%%%%%%%%%%%%%%%%%%%%%%%%%%%%%%%%%%%%%%%%%%%%
\section{Waveform accuracy for the Allen $\chi^2$ discriminator}
\label{s:WaveformAccuracy}
%%%%%%%%%%%%%%%%%%%%%%%%%%%%%%%%%%%%%%%%%%%%%%%%%%%%%%%%%%%%%%%%%%%%%%%%%%%%%%%

How do inaccuracies in approximate model waveforms affect the value of
$\chi^2_r$ defined in Eq.~(\ref{e:ChiSquareDef})?  Let $\delta h
= \sum_k\delta h^k$, where $\delta h^k$ denotes the component of the
model waveform error in the $k^\mathrm{th}$ frequency bin.  We find it
helpful to re-express the quantity $\langle h_e|h^k_m-h_m/p\rangle$
that appears in the definition of $\chi^2_r$ in
Eq.~(\ref{e:ChiSquareDef}) in the following way:
\begin{eqnarray}
\langle h_e|h^k_m-h_m/p\rangle &=&\langle h_m-\delta h|h_m^k-h_m/p\rangle
\nonumber\\
&=& -\langle \delta h^k|h_m\rangle.
\end{eqnarray}
The derivation of this expression depends on using the model waveform
normalization conditions given in Eqs.~(\ref{e:WaveformNormalization})
and (\ref{e:WaveformPhaseMatching}).  Using this expression, it is
straightforward to determine how waveform errors $\delta h$ affect the
value of $\chi^2_r$:
\begin{eqnarray}
\delta\chi^2_r = \frac{p}{2p-2}\frac{1}{\rho_m^2}
\sum_{k=1}^p\Bigl|\langle \delta h^k|h_m\rangle
\Bigr|^2.
\end{eqnarray}
If we define
\begin{eqnarray}
\delta h_\parallel^k=  h_m^k \sqrt{\frac{p^2}{2p-2}}\,
\frac{\langle \delta h^k|h_m\rangle}
{\langle h_m|h_m\rangle},
\label{e:hparalleldef}
\end{eqnarray}
then our expression for $\delta\chi^2_r$ can be written even more simply:
\begin{eqnarray}
\delta\chi^2_r=
\sum_{k=1}^p \langle\delta h^k_\parallel | \delta h^k_\parallel\rangle.
\label{e:DeltaChiSquare}
\end{eqnarray}
We point out that Allen~\cite{Allen2005} derived an analogous
expression (i.e., his Eq. 6.18) for the variation of his original
$\chi^2$ due to errors in the model-waveform parameters $\lambda_m$.
The current definition of $\chi^2_r$ in Eq.~(\ref{e:ChiSquareDef})
differs from Allen's original in significant ways: Allen's original
$\chi^2$ only measured the frequency dependence of differences in the
amplitudes, but not differences in phase, between the observed and
model waveforms.  And Allen did not consider the possibility of
intrinsic waveform errors in his analysis.
Equation~(\ref{e:DeltaChiSquare}) is significantly more general than
Allen's expression, and is therefore essentially new.

A reasonable requirement on the accuracy of model waveforms used to
evaluate $\chi^2_r$ , is that $\delta\chi^2_r$ be smaller than typical
random variations in $\chi^2_r$ due to Gaussian noise in the detector,
i.e., $\delta\chi_r^2\le 1/\sqrt{p-1}$ from
Eq.~(\ref{e:ChiSquareRange}).  This requirement on the intrinsic model
waveform error is given by
\begin{eqnarray}
\sum_{k=1}^p \langle\delta h^k_\parallel | \delta h^k_\parallel\rangle
< \frac{1}{\sqrt{p-1}}.\label{e:ChiSquareVetoRequirement} 
\end{eqnarray}
This expression makes it clear that the Allen $\chi^2$
discriminator imposes different accuracy requirements than those
needed for detection: Equation~(\ref{e:ChiSquareVetoRequirement})
places restrictions on $\delta h_\parallel^k$ instead of $\delta h$
itself.  Also, since the right side of
Eq.~(\ref{e:ChiSquareVetoRequirement}) is independent of the signal's
SNR, the {\it relative} waveform accuracy, $\delta h/h$, required by
Eq.~(\ref{e:ChiSquareVetoRequirement}) is more stringent for
higher-SNR signals.  This suggests that the model waveforms intended
for general use in gravitational wave data analysis should be tested
with respect to both of these requirements.

These waveform accuracy requirements, Eqs.~(\ref{e:DetectionAccuracy})
and (\ref{e:ChiSquareVetoRequirement}), can also be expressed in a
more intuitive way.  We define real quantities $\psi$ and $\varphi$
that represent the (log of the) amplitude and the phase of the
frequency domain waveforms respectively:
\begin{eqnarray}
h_e &=& e^{\psi_e + i \varphi_e},\\ 
h_m &=& e^{\psi_e+\delta \psi + i \varphi_e + i\delta \varphi}.
\end{eqnarray}
The waveform modeling error $\delta h=h_m-h_e$ can therefore be written
in the form:
\begin{eqnarray}
\delta h &=& h_e(e^{\delta\psi+i\delta\varphi}-1),\\
&\approx& h_m\Bigl[\delta\psi+i\delta\varphi 
+ {\mathcal O}(\delta h^2)\Bigr].
\label{e:AmplitudePhase}
\end{eqnarray}
(We assume that $|\delta h|\ll|h_m|$ and keep only the lowest order
terms in $\delta h$ in the following analysis.)  Using these
expressions, the left side of Eq.~(\ref{e:DetectionAccuracy}) (the
detection waveform accuracy requirement) becomes
\begin{eqnarray}
\frac{\langle \delta h | \delta h\rangle}{\langle h_m | h_m\rangle}
&=& \int_0^\infty \frac{4\delta h^*(f) \delta h(f) }
{S_n(f) \langle h_m| h_m \rangle}df,\\
&=&\int_0^\infty (\delta \psi^2+\delta\varphi^2)\, w(f) df,
\end{eqnarray}
where the weight function $w(f)$ is defined by
\begin{eqnarray}
w(f)&=& \frac{4\,h_m^*(f) h_m(f)}{S_n(f) \langle h_m | h_m \rangle}.
\end{eqnarray}
This signal-to-noise weighting function satisfies the usual
normalization condition $1=\int_0^\infty w(f)df$.  It will be useful
to denote $w(f)$-weighted averages of quantities in the following way,
\begin{eqnarray}
\overline{Q}\equiv \int_0^\infty Q(f)\,w(f)\,df.
\end{eqnarray}
Then Eq.~(\ref{e:DetectionAccuracy}) can be re-written in
terms of the amplitude and phase errors:
\begin{eqnarray}
\frac{\langle \delta h | \delta h\rangle}{\langle h_m | h_m\rangle}
=\overline{\delta\psi^2} +
\overline{\delta\varphi^2} \leq 2 \epsilon_{\mathrm{max}}.
\label{e:AlternateDetectionLimitII}
\end{eqnarray}

Next we want to express our Allen $\chi^2$-based model waveform
accuracy requirement, Eq.~(\ref{e:ChiSquareVetoRequirement}), in terms
of the waveform amplitude and phase errors.  Therefore we decompose
those waveform errors into amplitude and phase errors, $\delta \psi_k$
and $\delta \varphi_k$, having support in each frequency bin labeled
by the index $k$:
\begin{eqnarray}
\delta h^k = h_m(\delta \psi_k + i \delta\varphi_k).
\end{eqnarray}
The projection $\langle \delta h^k| h_m\rangle$ that
appears in the definition of $\delta h^k_\parallel$,
Eq.~(\ref{e:hparalleldef}), can be written in terms of the amplitude
and phase errors as
\begin{eqnarray}
\!\!\!\!\!\!\!\!\!\!\!
\frac{\langle \delta h^k| h_m\rangle}{\langle h_m|h_m\rangle}
&=& 4\int_0^\infty \frac{\delta h^k h_m^{*}}
{S_n\langle h_m|h_m\rangle}df,\\ 
&=& 4\int_0^\infty\frac{\bigl(\delta\psi_k 
+i\delta\varphi_k\bigr)h_m h^{*}_m }
{S_n\langle h_m|h_m\rangle}df,\\
&=& \overline{\delta\psi_k}
+i\overline{\delta\varphi_k}.
\end{eqnarray}
Using Eqs.~(\ref{e:hparalleldef}) and (\ref{e:DeltaChiSquare}), 
the effects of waveform error on $\delta \chi^2_r$
can therefore be written as
\begin{eqnarray}
\!\!\!\!\!\!\!\!\!
\delta\chi^2_r&=&
\frac{p\,\langle h_m|h_m\rangle}{2p-2}
\sum_{k=1}^p\left[\Bigl(\overline{\delta\psi_k}\Bigr)^2
+\Bigl(\overline{\delta\varphi_k}\Bigr)^2\right].
\label{e:ChiSquareAmplitudeExpression}
\end{eqnarray}
Equation~(\ref{e:ChiSquareVetoRequirement}) can
therefore be re-written as a requirement on the signal- and
detector-noise-weighted averages of the waveform amplitude and phase
errors:
\begin{eqnarray}
\!\!\!\!\!\!\!\!\!
\sum_{k=1}^p\left[\Bigl(\overline{\delta\psi_k}\Bigr)^2
+\Bigl(\overline{\delta\varphi_k}\Bigr)^2\right]
\leq
\frac{2\sqrt{p-1}}{p\,\langle h_m|h_m\rangle}.
\label{e:AlternativeChiSquareRequirement}
\end{eqnarray}
We note that the sums $\overline{\delta\psi} =\sum_k\overline{\delta
  \psi_k}$ and $\overline{\delta\varphi} =\sum_k\overline{\delta
  \varphi_k}$ vanish,
$\overline{\delta\psi}=\overline{\delta\varphi}=0$, as a consequence
of the waveform normalization conditions in
Eqs.~(\ref{e:WaveformNormalization}) and
(\ref{e:WaveformPhaseMatching}).

As an example, we examine these waveform accuracy requirements for
typical values of $p$ and $\epsilon_\mathrm{max}$ used in LIGO data
analysis: $p=16$ and $\epsilon_\mathrm{max}=0.005$.  In this case
Eqs.~(\ref{e:AlternateDetectionLimitII}) and
(\ref{e:AlternativeChiSquareRequirement}) reduce to
\begin{eqnarray}
\sqrt{\overline{\delta\psi^2} +
\overline{\delta\varphi^2}} &\lesssim& 0.1,
\label{e:DetectionIII}
\\
\sqrt{\sum_{k=1}^p\left[\Bigl(\overline{\delta\psi_k}\Bigr)^2
+\Bigl(\overline{\delta\varphi_k}\Bigr)^2\right]}
&\lesssim& 0.1\frac{7}{\sqrt{\langle h_m|h_m\rangle}}.\nonumber\\
\label{e:ChiSquareIII}
\end{eqnarray}
The right sides of these two inequalities are comparable for
$\rho_m=\sqrt{\langle h_m|h_m\rangle}=7$, but, as noted above, the
Allen $\chi^2$-based requirement becomes more restrictive for stronger
sources.

We recall that our Allen $\chi^2$-based waveform accuracy requirement
was derived assuming that the model waveform parameters have been
optimized to maximize $\rho_m^2$.  If the Allen $\chi^2$ discriminator
were to be used as a strict veto of candidate signals identified in a
template bank search with discretely spaced model waveform parameters,
then the effects of model waveform parameter mismatch would also have
to be taken into account.  This would likely decrease the model
waveform accuracy error tolerance for the Allen $\chi^2$
discriminator, as it does with the detection accuracy requirements
(cf. Ref.~\cite{Lindblom2008}).  However those new requirements would
depend critically on how this $\chi^2$ discriminator is used, e.g.,
how veto thresholds are set.  Since the Allen $\chi^2$ discriminator
is not presently being used in this way, however, we have forgone this
analysis here.

%%%%%%%%%%%%%%%%%%%%%%%%%%%%%%%%%%%%%%%%%%%%%%%%%%%%%%%%%%%%%%%%%%%%%%%%%%%
% Acknowledgment
%%%%%%%%%%%%%%%%%%%%%%%%%%%%%%%%%%%%%%%%%%%%%%%%%%%%%%%%%%%%%%%%%%%%%%%%%%%
\acknowledgments 
%%%%%%%%%%%%%%%%%%%%%%%%%%%%%%%%%%%%%%%%%%%%%%%%%%%%%%%%%%%%%%%%%%%%%%%%%%%
We thank Jolien Creighton, Benjamin Owen, and Xavier Siemens for
helpful comments on this work.  CC's research was supported in part by
grant PHY-1404569 to the California Institute of Technology from the
National Science Foundation.  LL's research was supported in part by
grants PHY-1604244 and DMS-1620366 to the University of California at
San Diego from the National Science Foundation.
\vfill
%%%%%%%%%%%%%%%%%%%%%%%%%%%%%%%%%%%%%%%%%%%%%%%%%%%%%%%%%%%%%%%%%%%%%%%%%%%%%%%
%References
%%%%%%%%%%%%%%%%%%%%%%%%%%%%%%%%%%%%%%%%%%%%%%%%%%%%%%%%%%%%%%%%%%%%%%%%%%%%%%%
\bibstyle{prd} \bibliography{../References/References}
%%%%%%%%%%%%%%%%%%%%%%%%%%%%%%%%%%%%%%%%%%%%%%%%%%%%%%%%%%%%%%%%%%%%%%%%%%%%%%%

\end{document}